\documentclass{aa}
\usepackage{graphics,epsfig,amsmath,amssymb,amstext}

\def\d {\mathrm{d}}
\def\Msun{\hbox{$M_{\odot}$}}

\begin{document}

\thesaurus{12(02.01.1; 02.14.1; 09.03.2; 10.01.1)}

\title{Superbubbles and the Galactic evolution of Li, Be and B}

\author{Etienne Parizot}

\institute{Dublin Institute for Advanced Studies, 5 Merrion Square, 
Dublin 2, Ireland; e-mail: parizot@cp.dias.ie}

\date{Received date; accepted date}

%\authorrunning{E. Parizot}
%\titlerunning{Superbubbles and the LiBeB evolution}
\maketitle

\begin{abstract}
	From a recent re-analysis of the available data, Be and B Galactic
	evolution appears to show evidence for a two-slope behaviour with
	respect to O. The inferred Be/O abundance ratio in halo stars is
	constant at very low metallicity (primary behaviour) and increases
	proportionally to O/H at high metallicity (secondary behaviour). 
	We show that this can be explained in the framework of one single
	model, the `superbubble model', in which Li, Be and B are produced
	by spallation reactions induced by energetic particles accelerated
	out of a mixture of supernova (SN) ejecta and ambient interstellar
	medium inside superbubbles (SBs).  All the qualitative and
	quantitative constraints, including the energetics, the value of
	the transition metallicity and the $^{6}$Li/$^{9}$Be isotopic
	ratio, are satisfied provided that the energetic particles have a
	spectrum flattened at low energy (in $E^{-1}$) and that the
	proportion of the SN ejecta inside SBs is of the order of a few
	percent.  This lends support to Bykov's acceleration mechanism
	inside SBs and to the SB dynamical evolution model of Mac Low \&
	McCray (1988).  
\end{abstract}

\section{Introduction}
\label{Introduction}

The issue of Galactic chemical evolution is one of the most
fundamental in astrophysics.  How does one pass from the state of the
universe following primordial nucleosynthesis, where virtually all the
baryonic matter is composed of H and He nuclei (plus about ten percent
of the current amount of $^{7}$Li), to a state in which the heavier
chemical elements (i.e. the `metals') exist in a sufficient proportion
to form planets, rocks, crystals, plants, animals and the bodies of
human beings?  Most of these elements are synthesized in stellar cores
or during the explosion of massive stars, and are released
progressively in the interstellar medium (ISM), increasing the
Galactic content in metals, i.e. the \emph{metallicity}, $Z$.  The
case of Li, Be and B (LiBeB), however, is different.  These
\emph{light elements} are believed to be produced through spallation
reactions induced by the interaction of energetic particles (EPs) in
the ISM. In these reactions, an heavier nucleus (most significantly C,
N or O) is `broken into pieces', or \emph{spalled}, and transmuted
into one of the lighter $^{6}$Li, $^{7}$Li, $^{9}$Be, $^{10}$B or
$^{11}$B nuclei.  Except for $^{7}$Li, spallative nucleosynthesis is
thought to be the main (if not the only) light element production
mechanism.  The case of $^{11}$B is also slightly more complicated, as
neutrino-induced spallation in supernovae (the so-called
$\nu$-process) is sometimes invoked to increase the B/Be and
$^{11}\mathrm{B}/^{10}\mathrm{B}$ ratios which one would expect should
the light elements be produced by nucleo-spallation alone.

Because of the spallative origin of LiBeB, the Galactic evolution of
light elements is also relevant to nuclear and high energy
astrophysics, as it provides unique constraints about the history of
EP interactions in our Galaxy.  While constraints about the
\emph{current} EP content of the Galaxy (i.e. cosmic rays as well as
possibly other components) can be derived from gamma-ray astronomy and
the study of ISM heating and ionization, information about the
\emph{past} EP populations can only be derived from integrated
observables, such as the abundance of LiBeB in stars of various
metallicities.  These abundances are the outcome of ongoing LiBeB
production from the `birth of the Galaxy' to the time when the
observed stars formed.  Therefore, one can regard low-metallicity
stars as astrophysical fossils giving evidence for the efficiency of
nucleosynthetical processes over time, or in other words for the pace
at which the chemical evolution of the various elements took place.

Assuming that the composition observed today at the surface of the
stars reflects that of the gas from which they formed, one can follow
the increase of the Be and B content of the Galaxy, say, as a function
of metallicity.  The inferred chemical evolution is thus not expressed
with respect to the usual physical time, but with respect to what we
can call a \emph{chemical time}, defined as the abundance of metals in
the ISM. In the approximation of a homogeneous Galaxy (without infall
of primordial gas), the chemical time is a mere redefinition and a
monotonically increasing function of the physical time.  If the Galaxy
is not chemically homogeneous, which is both expected and observed at
early stages, then the chemical time has to be defined locally and can
be a non monotonic function of ordinary time (e.g. if dilution of high
metallicity gas by low metallicity gas occurs; see Parizot \& Drury
1999b).

A further complication, however, arises from the fact that the ISM
metallicity proves to be a rather ambiguous measure of the chemical
time.  This is because the abundances of various elements are not
found to increase proportionally to one another, and in particular the
(most often used) O and Fe clocks do not appear to mark the same
unambiguous chemical time.  Indeed, according to some recent
measurements, the O/Fe abundance ratio is \emph{not} constant in the
Galaxy, even at low metallicity, but decreases as a function of O/H
(Israelian, et al.  1998; Boesgaard, et al.  1999; Mishenina, et al. 
2000).  This conclusion, however, is still controversial and there is
no general consensus about which of the available methods (OI IR
triplet, OI forbidden line, or OH molecular line) should be used to
derive the O abundance whenever they are conflicting (see however
Israelian et al.  2000, for a reconciliation between the various
methods).  It might therefore seem safer, when studying Be and B
evolution from stellar abundances, to use the Fe abundance as the
chemical time of reference, since it is in principle easier to
measure.  However it is O, not Fe, which is involved in the spallation
reactions producing LiBeB, hence in a way Fe is totally irrelevant to
the problem.  In the absence of a chemical clock both reliable and
relevant to light element Galactic evolution, we shall use here O as
the reference element, and discuss LiBeB Galactic evolution as a
function of O/H, keeping in mind that the exact numerical values might
be reconsidered (or confirmed!)  in the future, when a consensus about
the various observational methods is reached.

\section{Phenomenology and observational constraints}
\label{sec:phenomenology}

Among light elements the $^{7}$Li isotope is characterized by a
significant production by primordial nucleosynthesis, which accounts
for about 10\% of the total Li content of the Galaxy.  In comparison,
the contribution of Galactic nucleosynthesis (whether spallative or
from AGB stars or novae) is negligible in the early Galaxy, and the Li
stellar abundances are found to be almost constant at stellar
metallicities lower than about 1/20th solar (e.g. Spite \& Spite
1982).  This implies that the early Galactic chemical evolution of Li
is not constrained by the data, apart from being required to preserve
the observed Spite plateau.  In principle, this does not apply to
$^{6}$Li, which is not produced by primordial nucleosynthesis any
significantly.  However, $^{6}$Li represents such a tiny fraction of
the total Li present in a low-metallicity star that the measurement of
its specific abundance is a considerable observational challenge,
which has been taken up for only two halo stars up to now (Hobbs \&
Thorburn 1994, 1997; Smith et al. 1998; Cayrel et al. 1999).  Both
of these stars have a metallicity around $[\mathrm{Fe/H}] = -2.3$, and
show a $^{6}$Li abundance compatible with a $^{6}$Li/$^{9}$Be ratio
between 20 and 80, in contrast with the solar value of $\sim 6$.

As far as Be and B (BeB) are concerned, the contribution of primordial
nucleosynthesis is entirely negligible so that the abundances measured
in any low-metallicity star provide direct evidence about Galactic
nucleosynthesis itself.  From the phenomenological point of view, two
specific behaviours can be identified for Be and B Galactic evolution:
\begin{itemize}
	\item a \emph{primary} behaviour, in which the Be and B abundances
	increase proportionally to O/H, so that the Be/O and B/O abundance
	ratios are constant, and
    \item a \emph{secondary} behaviour, in which the Be and B
    abundances increase proportionally to the \emph{square} of the O
    abundance, so that the Be/O and B/O ratios are proportional to
    O/H.
\end{itemize}

The first case is expected if the Be and B production rate is
proportional to the rate of O release in the ISM, i.e. more or less to
the SN explosion rate, while the second case corresponds to the
standard scenario of Galactic cosmic ray nucleosynthesis (GCRN). 
Indeed, according to GCRN, Be and B are produced by spallation
reactions induced by the interaction of EPs accelerated out of the ISM
(namely, the cosmic rays) with the ISM itself (Reeves et al. 1970;
Meneguzzi et al. 1971).  The energy source of these EPs is the
kinetic energy of the SNe, so that the power available for spallation
is proportional to the SN explosion rate.  Now the effective
production of Be and B through spallation reactions depends on the
abundance of O among the EPs and the ISM (since O is the main
progenitor of Be and B).  The total production rate is thus
proportional not only to the rate of O release in the ISM, as in the
first case above, but also to the O abundance itself, i.e. our
chemical clock, O/H. This results in the secondary behaviour first
described by Vangioni-Flam et al.  (1990).

It has to be noted that the GCRN would also lead to a primary
behaviour, \emph{if the EP composition were constant and O-rich}. 
Indeed, most of the Be and B production would then be due to
interactions of energetic O nuclei with the ISM, and thus be
independent of the ambient O abundance.  The Be and B production rates
would then be simply proportional to the SN explosion rate, i.e. to
the rate of O release in the Galaxy.  One can thus conclude that,
phenomenologically, a primary process is expected when the O abundance
among the EPs is independent of (and higher than) that in the ISM, and
a secondary process is expected when the EP composition reflects that
of the ISM.

%%%%%%%%%%%%%%
\begin{figure}
\centerline{\psfig{file=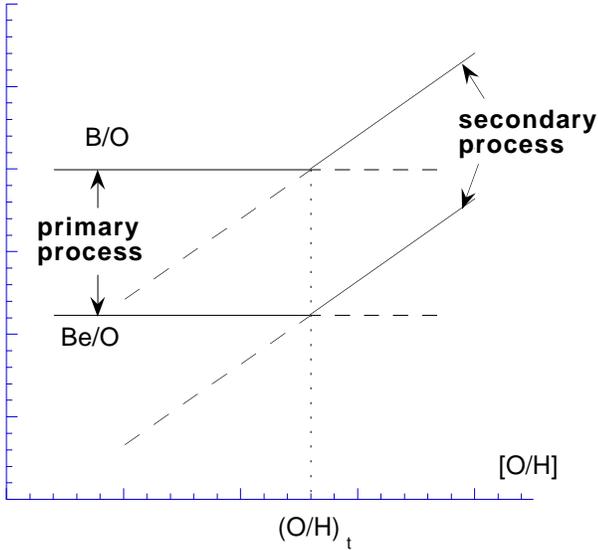,width=8 cm}}
\caption{Sketch of the expected Be and B evolution diagram in the case
when both a primary and a secondary process exist in the Galaxy.  The
elemental ratios Be/O and B/O are shown as a function of O/H, in
logarithmic scales.}
\label{phenomenology}
\end{figure}
%%%%%%%%%%%%%%

Now let us suppose that both a primary and a secondary process for BeB
production exist in the Galaxy.  Then for obvious reasons, and
whatever their respective weight in the global Galactic ecology, the
primary process is bound to dominated at very low metallicity, while
the secondary process must dominate in the high metallicity limit. 
This is simply because the efficiency of the (idealized) secondary
process is zero at $Z = 0$ and infinite at $Z \rightarrow \infty$. 
The abundance diagram showing Be/O and B/O as a function of O/H should
then look like in Fig.~\ref{phenomenology}, with a constant ratio
below some \emph{transition metallicity},
$Z_{\mathrm{t}}\equiv(O/H)_{\mathrm{t}}$, and a linearly increasing
ratio above.  Clearly, the precise determination of $Z_{\mathrm{t}}$
is an important goal for observational studies and would provide a
strong constraint on the theoretical models of Galactic Be and B
evolution.

In a recent study, Fields et al.  (2000) have re-analyzed the
available data about Be and B evolution as a function of O/H,
discussing the uncertainties associated with the various methods used
to derive the O abundance, the stellar parameters and the
incompleteness of the samples.  Their results are in conformity with
the picture proposed above.  In particular they have looked for
evidence of a transition metallicity, using statistical analysis, and
found a range of probable values of $\log(Z_{\mathrm{t}}/Z_{\odot})$
between $-1.9$ and $-1.4$ (see also Olive 2000).  Above this
metallicity, the Be and B evolution shows a secondary behaviour and
seems compatible with the standard GCRN scenario.  Unfortunately, very
few data points have yet been reported below $Z_{\mathrm{t}}$, and its
exact value is rather uncertain.  However, energetics arguments show
that a primary process \emph{is} indeed required below, say,
$10^{-2}Z_{\odot}$ (e.g. Parizot \& Drury 1999a,b; Ramaty et al.,
2000a,b).  This provides further support to the general `two-slopes'
scheme sketched out above, and in particular to the very existence of
a transition metallicity (which is also certified up to a 99\%
confidence level by Fields et al., 2000).  Since the energetics
argument has been somewhat controversial, we shall review in detail
how it works in Sect.~\ref{sec:energetics}.

Beforehand, let us summarize the available observational evidence as
follows:
\begin{enumerate}
    \item the Be/O ratio observed in halo stars is constant up to a 
    metallicity $Z_{\mathrm{t}}$,
	
	\item below $Z_{\mathrm{t}}$, $\mathrm{Be/O}\sim 4\,10^{-9}$ (see
	Sect.~\ref{sec:energetics}),
    
	\item $Z_{\mathrm{t}}$ is between $\sim 10^{-2}$ and
	$10^{-1.5}Z_{\odot}$,
    
	\item above $Z_{\mathrm{t}}$,  Be/O $\propto$ O/H,
	
	\item at $Z=Z_{\odot}$, Be/O $\simeq 3.1\,10^{-8}$ (Anders \&
	Grevesse 1989)
\end{enumerate}

To these must be added the constraints relating to Li:
\begin{enumerate}
	\addtocounter{enumi}{5}
	
	\item Li/Be $\la 100$ for $Z\la 10^{-1}Z_{\odot}$ (otherwise
	spallative nucleosynthesis breaks the Spite plateau),
	
	\item $^{6}$Li/$^{9}$Be is between 20 and 80 at $[\mathrm{Fe/H}]
	\simeq -2.3$, i.e. $[\mathrm{O/H}] \sim -1.5$ (with the observed
	O/Fe trend; Israelian et al., 2000), and
    
	\item $^{6}\mathrm{Li}/^{9}\mathrm{Be} \simeq 6$ at solar
	birth.
\end{enumerate}

Finally, the constraints relating to B read:
\begin{enumerate}
	\addtocounter{enumi}{8}
	
	\item $10 \le\mathrm{B}/\mathrm{Be} \le 30$;
	
	\item $^{11}\mathrm{B}/^{10}\mathrm{B} \simeq 4$ at solar birth;
\end{enumerate}

In the following, we show that the so-called superbubble model for
LiBeB production satisfies the whole of the above constraints with
only one free parameter plus an additional mechanism for B production,
e.g. the $\nu$-process or nucleo-spallation induced by a component of
low-energy cosmic rays (LECR).  This model has been shown by Parizot
\& Drury (1999b) to offer a natural and efficient primary process for
Be and B production in the very early Galaxy.  Here, we argue that it
also predicts the required change of behaviour (from primary to
secondary) at a transition metallicity in the range derived by Fields
et al.  (2000) from the observational data (this prediction was
actually implied by Fig.~3 of Parizot \& Drury 1999b, and the related
discussion, although not stated explicitly).  In other words, the
full range of $^{6}$LiBeB evolution (i.e. below \emph{and} above the
primary-to-secondary transition) can be accounted for by one single
mechanism, namely the acceleration of particles inside a superbubble
(SB) out of a mixture of SN ejecta and ambient ISM.

In the following section, we review the GCRN energetics.  The SB model
is presented in Sect.~\ref{sec:SBmodel}, and the ingredients of the
calculation are derived in Sect.~\ref{sec:ingredients}.  These
ingredients are the composition and the energy spectrum of the
superbubble energetic particles (SBEPs).  In Sect.~\ref{sec:results},
we present the results relating to the evolution of the LiBeB
abundances as a function of O/H. We discuss their implications in
Sect.~\ref{sec:conclusion}.

\section{The energetics of GCRN}
\label{sec:energetics}

%%%%%%%%%%%%%%
\begin{figure}
\centerline{\psfig{file=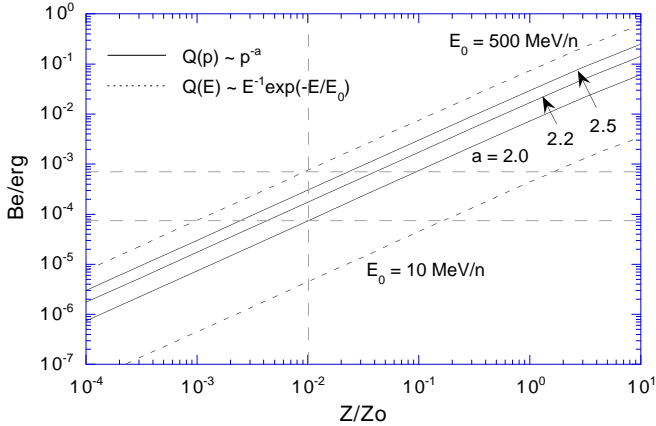,width=\columnwidth}}
\caption{Be production efficiency by spallation for different
energy spectra, in numbers of Be nuclei produced per erg injected, as a
function of the ISM (target) metallicity.}
\label{BeProdEff}
\end{figure}
%%%%%%%%%%%%%%

In this section, we intend to show that, whatever the assumption about
the O/Fe trend in halo stars, the standard GCR nucleosynthesis cannot
account for the Be abundance in low metallicity stars, because of
energetics problems.

In their analysis, Fields et al.  (2000) consider three sets of Be, B
and O data (which they call the Balmer data, the IRFM1 data, and the
IRFM2 data), differing by the choice of stellar parameters used to
derive the abundances (see also Fields \& Olive 1999).  For the first
two data sets, the resulting Be abundance found at
$[\mathrm{O}/\mathrm{H}] = -2$ (i.e. $\mathrm{O}/\mathrm{H} = 10^{-2}
(\mathrm{O}/\mathrm{H})_{\odot})$ is consistent and equal to
$\mathrm{Be}/\mathrm{H} = 2\,10^{-14}$.  For the third data set,
namely IRFM2, the Be abundance found at the same metallicity is
somewhat higher: $\mathrm{Be}/\mathrm{H} \sim 3\,10^{-14}$ as derived
from Fields' best model, and $\mathrm{Be}/\mathrm{H} \sim 5\,10^{-14}$
as derived directly from the data.  In order to obtain a compelling
conclusion, we use here the most conservative (i.e. lowest) value of
$\mathrm{Be}/\mathrm{H} = 2\,10^{-14}$ at $[\mathrm{O}/\mathrm{H}] =
-2$, i.e. $\mathrm{O}/\mathrm{H} = 5\,10^{-6}$ (Anders \& Grevesse,
1989).  The Be/O ratio is then $4\,10^{-9}$, which means that, on
average, once integrated over the Galactic evolution up to the time
when $[\mathrm{O}/\mathrm{H}]$ reaches $-2$, the production of each
nucleus of O must has been accompanied with the production of
$4\,10^{-9}$ nuclei of Be.  Considering that a SN produces on average
$\sim 1.5\,\Msun$ of oxygen, i.e. $\sim 1.1\,10^{56}$ O nuclei, we end
up with an average required production of Be of $\sim 4.5\,10^{47}$
nuclei per supernova (to be scaled linearly with the actual O yield
one wishes to adopt for the SNe).

According to GCRN (and most models of Be production), the EPs
responsible for the CNO spallation draw their energy from SN
explosions.  Therefore, assuming that about 10\% of the SN energy is
eventually imparted to the EPs, the total energy available for
spallation is of order $10^{50}$ erg, which implies a Be production
efficiency of $\sim 4.5\,10^{-3}$ nuclei synthesized per erg of EPs
injected in the ISM. This efficiency has to be compared with the
theoretical expectations.  We have calculated the Be production
efficiency as a function of the target metallicity for different EP
energy spectra.  The EP composition has been taken identical to the
ISM composition, as appropriate for the GCRN scenario.  The results
are shown in Fig.~\ref{BeProdEff}.  It should be noted that an
increase of the metal abundances in the EPs of about a factor of 10,
as could be invoked to conform to the current over-metallicity of the
GCRs as compared to the mean ISM composition, does not change the
results by more than a factor of 1.7 (upwards).  As can be seen from
Fig.~\ref{BeProdEff}, the Be production efficiency obtained for the
GCRN, i.e. with the standard cosmic ray source spectrum in $p^{-2}$,
is only $\sim 7.5\,10^{-5}$ Be/erg, namely a factor 60 below the
required value (or even more if one considers that the value of
$4.5\,10^{-3}$ derived above is actually the mean value of the Be
production efficiency for $0\le Z_{\mathrm{ISM}}\le
10^{-2}Z_{\odot}$).

Allowing for a steeper slope of the cosmic ray source spectrum, namely
$p^{-2.2}$ or even $p^{-2.5}$, does improve the situation, but not
enough to reconcile the GCRN expectations with the observed amount of
Be in low metallicity stars.  Even with the most efficient spectrum,
$Q(E)\propto E^{-1}\exp[-E/(500~\mathrm{MeV/n})]$ (see
Sect.~\ref{SBEPEnergySpectrum}), the Be production efficiency obtained
is still a factor 7 below the required value at
$[\mathrm{O}/\mathrm{H}] = -2$, not mentioning the values at lower
metallicity.  We conclude that the standard GCRN cannot account for
the production of Be and B in the early Galaxy, say for $Z\le
10^{-2}Z_{\odot}$.

\section{The superbubble model}
\label{sec:SBmodel}

As emphasized in Parizot \& Drury (1999a,b), a simple solution to the
above energetics problem consists in letting more metal-rich EPs
interact with the ISM, so that the energetic carbon, nitrogen and
oxygen nuclei (CNO) can be spalled in flight while propagating through
the ISM and thereby increase the BeB production usually resulting
mainly from energetic protons and $\alpha$ particles interacting with
the ambient CNO. If the abundance of CNO among the EPs is large
enough, these \emph{reverse spallation} reactions can actually
dominate the BeB production in the ISM and make the production
efficiency independent of the ambient metallicity, in contrast with
the situation shown in Fig.~\ref{BeProdEff}.  This is typical of a
primary production mechanism, which has been shown to be bound to
dominate at very low metallicity.  Since the CNO nuclei in the ISM
originate from SN explosions, EPs with a high abundance of CNO can be
obtained if particles are accelerated from a material contaminated by
large amounts of SN ejecta.  This is the case inside the superbubbles
(SBs) which form consequently to the explosion of many SNe in an OB
associations.  Indeed, the dynamical effect of repeated SN explosions
in a small region of the Galaxy is to blow large (super)bubbles of
hot, rarefied material ($T > 10^{6}$K, $n \sim
10^{-2}\,\mathrm{g\,cm}^{-2}$), surrounded by shells of swept-up and
compressed ISM. The interior of superbubbles consists of the ejecta
and stellar winds of evolved massive stars \emph{plus} a given amount
of ambient ISM evaporated off the shell and dense clumps passing
through the bubble.

The exact composition of the particles to be accelerated inside the
bubble depends on i) how much ambient material has evaporated towards
the SB interior, ii) how well it is mixed with the SN ejecta, and iii)
where the acceleration occurs, whether over the whole SB or more
locally around the most recent SN explosion.  If the material inside
the superbubble is not well mixed, it is possible that the SBEPs be
richer (or poorer) in CNO than the average material, depending on
where exactly the acceleration takes place.  None of the three
questions above has yet received a conclusive answer, neither from
theoretical nor observational studies.  This explains why the
superbubble models proposed so far to account for Be and B Galactic
evolution (Parizot \& Drury 1999b,2000; Ramaty, et al., 2000a,b;
Bykov, et al., 2000) make different assumptions relative to the SB
dynamics and to the acceleration process.  In particular, the mass
fraction of the SN ejecta inside the superbubble is of the order of a
few percent if a thermal conduction model is used to evaluate the
evaporation of the material from the supershell, while it would be
much higher if the so-called `magnetic suppression' mechanism occurred
(Higdon et al. 1998).  Likewise, the mixing of the ejecta with the
ambient gas inside the bubble has been found to be efficient by
Parizot \& Drury (1999b), who compared the turbulent mixing time with
the age of the SB, while Ramaty et al.  implicitly assumed a
relatively poor mixing.  Finally, the acceleration mechanism has been
assumed by Parizot \& Drury to be of the Bykov type (Bykov \&
Fleishman 1992; Bykov 1995,1999), and thus more or less distributed
over the whole SB, while Ramaty et al.  use the standard shock
acceleration model, and thus restrict the acceleration process to the
vicinity of the latest SN explosion.  Note that different EP energy
spectra also result from these different assumptions (see below).

As indicated above, the current knowledge about superbubble dynamics
does not allow one to decide between the models and to determine
unequivocally the composition and spectrum of the SBEPs.  From the
point of view of nucleosynthesis, though, the SBEP composition and
spectrum are the only ingredients we need to calculate the production
rates of Li, Be and B and deduce their Galactic chemical evolution. 
Therefore, we propose here to proceed the other way round: first
investigate the validity of the superbubble scheme for the LiBeB
evolution, and second, use the available LiBeB data to constrain the
SB models.  To achieve the first point, we must show that the SB
models naturally produce a Be evolution scheme in agreement with the
available data, which applies to the qualitative \emph{and}
quantitative features.  In other words, the SB models have to account
for the total amount of Be found in the halo stars \emph{and} the
change of behaviour from a primary to a secondary evolution (with
respect to oxygen) around a transition metallicity $Z_{\mathrm{t}}$ in
the range currently allowed by the data.  Then we shall analyze more
precisely the link between $Z_{\mathrm{t}}$ and the parameters of the
SB so as to be able to draw sensible conclusions once $Z_{\mathrm{t}}$
is obtained with a greater precision from the data, hopefully in the
near future.  To this purpose, we now discuss the parameterization of
the SB model and derive the relevant physical ingredients (for more
details about the SB model itself, see Parizot \& Drury 1999b,2000).

\section{The physical ingredients of the model}
\label{sec:ingredients}

\subsection{The SBEP composition}

Concerning the SBEP composition, we assume that the material
accelerated inside the superbubble consists of a fraction $x$ of pure
SN ejecta, and a fraction $(1-x)$ of the ambient material, i.e. gas
with the ISM metallicity.  As the general (or average) metallicity of
the Galaxy increases, this ambient gas also gets richer and richer in
heavy elements.  Noting $\alpha_{\mathrm{ej}}(\mathrm{X})$ and
$\alpha_{\mathrm{ISM}}(\mathrm{X})$ the abundances of element X among
the SN ejecta and in the ISM, respectively, we can write the abundance
of X among the SBEPs as:
\begin{equation}
	\alpha_{\mathrm{SBEP}}(\mathrm{X}) = x\alpha_{\mathrm{ej}}(\mathrm{X})
	+ (1 - x)\alpha_{\mathrm{ISM}}(\mathrm{X}).
	\label{SBEPAbundances}
\end{equation}
Physically, the single parameter $x$ gathers all the information
relative to the amount of evaporation off the SB shell and the dense
clumps inside the SB (which governs the mass load of the SB by
interstellar gas) and the effective mixing of the evaporated gas with
the SN ejecta, before acceleration.  The model obtained with $x=1$
corresponds to the acceleration of pure ejecta, while $x=0$
corresponds to the standard GCRN (acceleration of uncontaminated ISM).

In principle, the decomposition of the SBEPs into these two components
(ejecta and ISM) can always be conceived in the abstract.  Therefore
the above parameterization does not introduce any simplification, as
long as the acceleration mechanism is assumed to be non-selective
(i.e. each chemical element is accelerated proportionally to its
abundance).  However, the value of the parameter $x$ should be
expected to vary from one SB to another, and to vary with time for
each given SB, due to its dynamical evolution (Parizot \& Drury,
1999b).  As we show below, different values of $x$ result in different
SBEP composition and thus different LiBeB yields.  The fact that $x$
can be different for different SBs in the Galaxy therefore results in
some scatter in the $^{6}$LiBeB data.  Additional causes for scatter
in the data have been analyzed in detail in Parizot \& Drury (2000)
and we shall not consider these features here.  Instead, we
concentrate on the mean evolution of Be and B in the Galaxy and
average our results over all the SBs participating to the $^{6}$LiBeB
enrichment.  This amounts to using a single value of $x$ in our
models, thought of as the average of the individual $x$'s of
individual SBs.  In the same manner, we neglect the time-dependence of
$x$ resulting from the SB dynamical evolution (short time scale, $\sim
10^{7}$ yr).  While the weight of the swept-up, evaporated ISM
increases inside the bubble as time passes and more and more SNe
explode, we use here an average $x$ over the whole SB evolution.  For
results and discussion of the full time-dependent treatment, see
Parizot \& Drury (1999b).  We also average over possible changes in
the Galactic physical conditions, like for instance the mean ISM
density and magnetic field (longer time scale, $\sim 10^{8}$--$10^{9}$
yr).

With the above parameterization, all we need to know to calculate the
SBEP composition as a function of time (or mean ISM metallicity) is
the composition of the ejecta and that of the ISM. The latter is
assumed to be the solar composition (taken from Anders \& Grevesse,
1989), with all the metals scaled in the same way.  Concerning helium,
we follow Ramaty et al.  (1997) and adopt a slightly decreasing
abundance with decreasing metallicity, to account for the He Galactic
enrichment, from the primordial to the solar value.  The function
which we use is slightly different from that of Ramaty et al., and is
given by:
\begin{equation}
	f_{\mathrm{He}}(Z_{\mathrm{ISM}}) \equiv
	\frac{\alpha_{\mathrm{ISM}}(\mathrm{He})}{\alpha_{\odot}(\mathrm{He})}
	= 1 + 0.0625\log\left(\frac{Z_{\mathrm{ISM}}}{Z_{\odot}}\right).
	\label{fHe}
\end{equation}
Note also that we scale the CNO abundances with O/H, rather than Fe/H.
According to the recent O/Fe observations discussed in
Sect.~\ref{Introduction}, the Fe abundance should not be scaled in the
same way.  However, this is not relevant to our calculations here
because Fe does not significantly contribute to the LiBeB production. 
Table~\ref{ISM(Z)} summarizes our prescription for the ISM metallicity
as a function of $[\mathrm{O}/\mathrm{H}]$.

\begin{table}
\caption[]{\label{ISM(Z)} ISM abundances relevant to Be and B
production, scaled with $Z_{\mathrm{ISM}}\equiv$ O/H
($f_{\mathrm{He}}$ is given by Eq.~\ref{fHe}).}
\begin{flushleft}
\begin{tabular}{cc}
\hline
Isotope & Abundance\\
\hline
 $^{1}$H  & $1.18\,10^{3}/10^{[\mathrm{O}/\mathrm{H}]}$ \\
 $^{4}$He & $1.18\,10^{2}\times f_{\mathrm{He}}/10^{[\mathrm{O}/\mathrm{H}]}$ \\
$^{12}$C  & $0.427$ \\
% $^{13}$C  & $4.82\,10^{-3}$ \\
$^{14}$N  & $0.132$ \\
$^{16}$O  & $1.00$ \\
\hline
\end{tabular}
\end{flushleft}
\end{table}

Finally, we need to specify the composition of the SN ejecta released
in the interior of the superbubble.  We use the models of Woosley \&
Weaver (1995).  The exact composition depends on the mass of the SN
progenitor and the explosion model used.  However, since we wish to
get information from the ensemble of the $^{6}$LiBeB data, we use a
mean composition obtained by averaging the yields of individual SNe
over a Salpeter IMF (initial mass function).  The results are shown in
Table~\ref{SNYields} for the relevant nuclei, as a function of the
initial metallicity of the progenitor.  It has to be noted that
Woosley \& Weaver (1995) used a Galactic chemical evolution model to
generate abundances appropriate to every metallicity investigated. 
Although this model is not specified in the paper, it most probably
does not take into account the O/Fe trend discussed above (which is
incompatible with the SN explosion models, unless a non-standard
chemical evolution model is used; see e.g. Ramaty, et al., 2000a,b). 
For this reason, the appropriate initial abundances should be
different from those used in the explosion simulations.  In the
absence of any reliable prescription to modify the results
accordingly, we use the published yields and assume that the quoted
metallicity refers to the oxygen abundance (which is not really the
case).  Since the average yields appear to be almost independent of
the initial metallicity for the nuclei of interest (H, He, C and O),
the error introduced by being mistaken in the actual initial
metallicities is expected to be negligible.  Although the nitrogen
yields are strongly dependent on the initial metallicity, the
influence of this element on the total production of Be and B has been
found marginal in any of our models.

\begin{table}
\caption[]{\label{SNYields} SN yields (in $\mathrm{M}_{\odot}$) of the
elements relevant to Be and B production, as a function of the initial
metallicity of the SN progenitor.  The ejected masses of individual
SNe have been taken from Woosley \& Weaver (1995), models A, and
averaged over a Salpeter IMF.}
\begin{flushleft}
\begin{tabular}{ccccc}
\hline
$Z_{\mathrm{ISM}}/Z_{\odot}$ & 0.0001 & 0.01 & 0.1 & 1 \\
\hline
 $^{1}$H  & 9.0 & 8.8 & 8.8 & 8.0 \\
 $^{4}$He & 6.6 & 6.7 & 6.8 & 6.8 \\
$^{12}$C  & 0.19 & 0.20 & 0.20 & 0.21 \\
$^{14}$N  & 1.5e-5 & 7.1e-4 & 6.6e-3 & 6.6e-2 \\
$^{16}$O  & 1.1 & 1.4 & 1.5 & 1.5 \\
\hline
\end{tabular}
\end{flushleft}
\end{table}

From the SN yields of Table~\ref{SNYields}, we obtain the elemental
abundances $\alpha_{\mathrm{ej}}(\mathrm{X})$ to be used in
Eq.~(\ref{SBEPAbundances}), and derive the abundances at intermediate
metallicities by fitting the `data points' linearly in logarithmic
scale (power law).  For completeness, we now give these fits for the
number abundances of H, He, C and N, normalized to
$\alpha_{\mathrm{ej}}(^{16}\mathrm{O}) = 1$.  Writing the abundance of
isotope $\mathrm{X}$ as $\alpha_{\mathrm{ej}}(\mathrm{X}) =
A\times(Z_{\mathrm{ISM}}/Z_{\odot})^{B}$, we obtain $(A,B) =
(83.139,0.048467)$ for $^{1}$H, $(17.062,0.034779)$ for $^{4}$He,
$(0.17435,0.027322)$ for $^{12}$C, and $(0.040795,0.87048)$ for
$^{14}$N, from which we see that the abundances of H, He, C and O are
almost constant in the SN ejecta, while the abundance of $^{14}$N
increases almost linearly with metallicity, as required for a
secondary nucleus.

\subsection{The SBEP energy spectrum}
\label{SBEPEnergySpectrum}

Concerning the SBEP energy spectrum, we investigate different shapes
which we discuss below:
\begin{enumerate}
	\item the standard cosmic ray source spectrum, $Q(p)\propto p^{-a}$,
	with $a = 2.0$, referred to as the `CRS2.0 spectrum';
	
	\item a modified (steeper) cosmic ray source spectrum,
	$Q(p)\propto p^{-a}$, with $a=2.2$ or even $a=2.5$, referred to as
	`CRS2.2' or `CRS2.5';
	
	\item the so-called `weak SB spectra', $Q(E)\propto
	E^{-1}\times\exp(-E/E_{0})$, designed to mimic the results of
	Bykov's acceleration mechanism in the case when most of the shocks
	responsible for particle acceleration inside the SB are weak
	shocks.
	
	\item the so-called `strong SB spectrum', matching the weak SB
	spectra below some break energy, $E_{\mathrm{break}}$,
	$Q(E)\propto E^{-1}$ or equivalently $Q(p)\propto p^{-1}$, and the
	CRS2.0 spectrum above $E_{\mathrm{break}}$, $Q(p)\propto p^{-2}$.
\end{enumerate}

The first spectrum, `CRS2.0', is motivated by the results of single
shock acceleration models.  We assume that it extends from 10~keV/n to
$10^{5}$~GeV/n.  Although a higher cut-off is possible, our results
only depend logarithmically on its actual value, through the amount of
energy stored (and thus lost) in essentially unproductive high energy
particles.  The second spectrum, `CRS2.2', is sometimes inferred for
the cosmic-ray source spectrum by propagating backwards the observed
GCR spectrum, assuming a plausible propagation model for the CRs in
the ISM (e.g. Engelmann et al. 1990).  As for the spectrum `CRS2.5',
it may be regarded as an \emph{extreme} cosmic-ray source spectrum,
and serves mainly as a `toy spectrum' here, for investigation
purposes.

The weak SB spectra have been used in previous works without clear
physical justification.  Their shape can be understood as follows. 
Considering that from the point of view of acceleration, the interior
of a superbubble behaves as an ensemble of thousands of shocks with
various velocities stochastically distributed over the SB volume
(Bykov \& Fleishman 1992), one can regard the SB acceleration model
as a variant of multiple shock acceleration already investigated by
several authors (see e.g. Marcowith \& Kirk 1999, and references
therein).  In these models, the low-energy particles are accelerated
very efficiently because their probability to escape from the shocks
is very low: whenever a particle is advected away from one particular
shock front, it can be `caught' again by and flow through another
shock in a different place.  Remembering the general relation between
the power-law index of particles accelerated at a single shock front
and the escape and acceleration timescales: $f(p)\propto p^{-s}$, with
$s = 3 + t_{\mathrm{acc}}/t_{\mathrm{esc}}$ (Kirk, et al. 1994), one
sees that as the effective escape time tends to infinity in a multiple
shock configuration, the power-law index of the EP distribution
function tends towards $3$.  Now a distribution function $f(p)\propto
p^{-3}$ is equivalent to an EP source spectrum in $Q(p) = f(p)\times
4\pi p^{2}\propto p^{-1}$ in our notations, or $Q(E) = Q(p)(\d p/\d
E)\propto E^{-1}$ in the non relativistic regime.  This explains the
shape of the weak SB spectra below the cut-off energy $E_{0}$.

Clearly this spectral shape cannot hold up to the highest energies,
since the total energy involved in EPs would then be strongly
divergent above $E \sim m_{\mathrm{p}}c^{2}$.  In fact, a cut-off is
expected in multiple shock acceleration models around an energy
$E_{0}$ such that the diffusion length of the particles having this
energy is comparable to the inter-shock distance.  For energies
greater than $E_{0}$, the assumptions normally used to derive the
basic transport equation are violated, and the above arguments fail. 
A comparable change in the acceleration regime around this energy is
also discussed by Bykov et al., and found to be responsible for the
steepening of the particle distribution function.  Following very
crude arguments, one could expect the spectrum above $E_{0}$ to
resemble the single-shock spectrum, $f(p)\propto p^{-s}$, where $s =
3r/(r - 1)$ and $r$ is the compression ratio across the shock
discontinuity.  Now most of the shocks are expected to be quite
\emph{weak} inside a superbubble (hence the name of the spectra),
since they are mostly secondary shock, created by the reflection of a
few primary shocks (from the SNe themselves or the stellar winds) over
clumps of denser material or other strong or weak shocks.  The
compression ratios across these weak shocks are thus probably less
than 2 and in fact possibly close to 1, which implies that the
power-law index $s$ is quite high.  Interestingly enough, as long as
$s$ is greater than 4, which corresponds to the standard CRS2.0
spectrum and is obtained only for strong shock ($r = 4$), its exact
value is not relevant to the calculations of LiBeB production because
only a negligible amount of particles have energies above the energy
$E_{0}$ around which the spectral shape changes.  The bulk of the
spallative LiBeB production inside the superbubble or in the
surrounding shell is thus due to the interactions of SBEPs with
energies below $E_{0}$, so that the actual shape of the cut-off is
unimportant.  This is the reason why we adopt an exponential cut-off
for the weak SB spectra instead of a power-law cut-off, in order to
avoid an extra free parameter, and keep only $E_{0}$ as the relevant
free parameter.  According to Bykov \& Fleishman (1992), $E_{0}$ is of
the order of a few hundreds or a few thousands of MeV/n.  We use here
500~MeV/n as a `canonical' value for the cut-off energy, but we also
explored other values and found no significant changes (except for
$E_{0} < 100$ MeV/n).

Finally, the `strong SB spectrum' can be thought of as a SB spectrum
in which the high energy particle distribution would match the
standard cosmic-ray source spectrum.  According to the above
discussion, this would occur if most of the shocks accelerating the
particles inside superbubbles were actually \emph{strong} shocks
(hence the name of the spectrum), with a compression ratio close to 4. 
In that case, the SBEPs could be but the GCRs observed at Earth.  This
`unifying scheme' actually corresponds to the original ideas of Bykov
\& Fleishman (1992), and has already been proposed in the context of
CR acceleration and LiBeB Galactic evolution (Higdon, et al. 1998;
Ramaty \& Lingenfelter 1999; Ramaty, et al., 2000a).  In addition,
it does not violate any available observation as the cosmic ray
spectrum at low energy is not observable at Earth because of the solar
modulation.  We would then simply be able to predict a flattening of
the GCR spectrum at low energy, approaching $Q(E)\propto E^{-1}$. 
Interestingly enough, we find that the `weak' and `strong' SB spectra
give very similar results as far as the LiBeB production is concerned. 
For this reason, we may loosely refer below to any of the SB spectra
as `the SB spectrum'.  And since the LiBeB data do not constrain the
SBEP spectrum at high energy (say above 1~GeV/n), both of the
following views are possible: 1) the SBEPs have a spectrum extending
up to energies of, say, $10^{5}$ GeV/n, and are just the cosmic rays
observed at Earth; or 2) the SBEP spectrum is cut at relatively low
energy, so that the SBEPs cannot be the same EP component as the GCRs.

% From this point of view, it is probably better to keep to the
% simplest assumption (involving only one EP component) and adopt the
% first position, as suggested by Ramaty et al.

It is worth noting, however, that all the spectra described above have
different implications for gamma-ray line astronomy as well as for the
ionization and heating of the gas in the supershells and the adjacent
molecular clouds.  This means that additional constraints on the SBEP
energy spectrum can be derived in principle from the calculation of
the nuclear excitation rates and the ionization rate resulting from
the interaction of the SBEPs with the ambient matter.  These questions
will be addressed elsewhere.  In this paper, we focus on the
implications of the different spectra for the LiBeB nucleosynthesis
only, without any prejudice coming from other fields of astrophysics,
to see whether the SB model can indeed account for the LiBeB
observational data and what constraints can be derived from them.  The
ultimate goal, of course, will be to gather and combine all the
possible constraints from various fields and improve our understanding
of the acceleration mechanisms inside a superbubble and of the SBEP
energy distribution.

\section{The LiBeB production induced by the SBEPs}
\label{sec:results}

%%%%%%%%%%%%%%
\begin{figure}
\centerline{\psfig{file=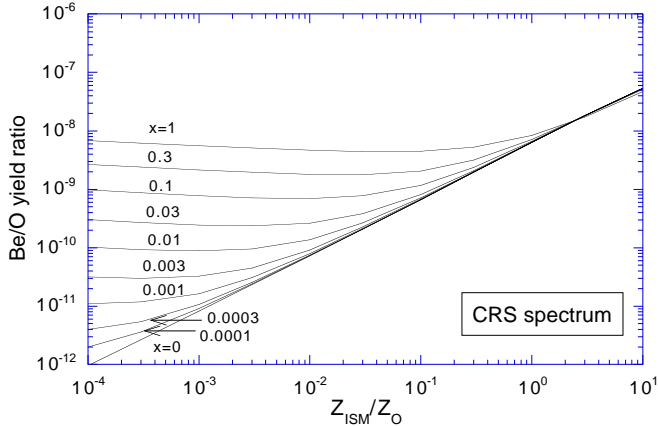,width=\columnwidth}}
\caption{Be/O yield ratio obtained with the CRS2.0 spectrum, as a
function of the ambient metallicity, for various values of the mixing
parameter, $x$.  The Be yield is calculated for a total energy of
$10^{50}$ erg imparted to the SBEPs.}
\label{Be/O(CRS2.0)}
\end{figure}
%%%%%%%%%%%%%%

We now have all the ingredients necessary to calculate the Li, Be and
B production induced by the interaction of the SBEPs with the ambient
matter.  While some of these interactions occur inside the
superbubbles themselves, the density there is so low that the
corresponding LiBeB production is negligible as compared to that
occurring in the dense regions surrounding the bubbles, namely the
supershells and part of the molecular clouds from which the parent OB
association formed.  Dense clumps trapped inside the SBs can also
provide efficient targets for the SBEPs.  In all cases, the chemical
composition of the target material is very close to the mean ISM
composition at the time when the SBEP acceleration occurs.  This
allows us to close our problem, since we have already calculated the
ISM composition as a function of $Z_{\mathrm{ISM}}/Z_{\odot}$ (see
Table~\ref{ISM(Z)}).  Knowing the composition of the projectiles,
their energy spectrum and the composition of the target, we can indeed
calculate the various reaction rates by using the model described in
Parizot \& Lehoucq (1999).  The relevant physical ingredients here
are:

\begin{enumerate}
	\item the spallation cross sections, taken from Read \& Viola
	(1984) and various updates summarized in Ramaty et al.  (1997);
	
	\item the total inelastic cross sections from which we determine
	the probability for the projectiles to be destroyed before they
	give rise to a spallation reaction, and for the daughter nuclei
	(Li, Be or B) to be destroyed before they can thermalize and be
	`integrated' to the ISM, so as to participate to the next episode
	of star formation (we use the semi-empirical values given by
	Silberberg \& Tsao 1990);
	
	\item the energy loss rates for the various nuclei in the medium
	of propagation, which compete with the nuclear reactions producing
	LiBeB. They have been kindly provided by J\"urgen Kiener.
\end{enumerate}

In order to compare the LiBeB production obtained from our SB models
with the observational data, we first study the absolute Be
production, as a function of the ISM metallicity, and then discuss the
elemental and isotopic ratios of the three light elements.

\subsection{Beryllium}

To derive the Galactic evolution of beryllium, we compute the Be yield
corresponding to the explosion of one supernova, and divide it by the
yield of oxygen at the same metallicity, as interpolated
logarithmically from Table~\ref{SNYields}.  This allows us to obtain a
Be/O yield ratio, which indicates how the Be content of the Galaxy
evolves as a function of the metallicity, $[\mathrm{O}/\mathrm{H}]$. 
We have assumed that a total of $10^{50}$ erg per SN is imparted to
the SBEPs, which amounts to approximately $10\%$ of the SN kinetic
energy.  Such a value for the particle acceleration efficiency is
quite typical of most acceleration models, but a simple scaling of our
results is straightforward for more (or less) efficient acceleration
mechanisms.

In Fig.~\ref{Be/O(CRS2.0)}, we show the Be/O yield ratio obtained for
the CRS2.0 spectrum and various values of the dilution parameter, $x$,
as a function of the ISM metallicity.  It is obvious from these curves
that each particular SB model, i.e. each particular value of $x$,
leads to a Be evolution which is primary ($\mathrm{Be}/\mathrm{O}
\sim$ constant) below some transition metallicity, $Z_{\mathrm{t}}$,
and secondary ($\mathrm{Be}/\mathrm{O} \propto \mathrm{O}$) above this
metallicity, as required by the data.  These qualitative features can
be understood as follows.  When the ambient metallicity is low, the
metallicity of the SBEPs is dominated by the SN ejecta, except for
very low values of the mixing parameter, $x$.  This is easily seen
from Eq.~(\ref{SBEPAbundances}): when $Z_{\mathrm{ISM}} =
10^{-4}Z_{\odot}$, say, the abundance of C, N and O is $\ga 10^{4}$
times lower in the ISM than in the ejecta; therefore, most of the CNO
present among the SBEPs come from the SN ejecta if $x \ga 10^{-4}$. 
As a result, there is more CNO in the SBEPs than in the ISM and the Be
production is dominated by reverse spallation reactions.  For
relatively large values of $x$, thus, the Be production efficiency is
independent of the ambient metallicity, and the Be/O yield ratio is
approximately constant (we even find slightly decreasing values
because the relative weight of H and He in the SBEPs varies a bit and
because the O yield itself is slightly increasing with
$Z_{\mathrm{ISM}}$).  When the ISM metallicity increases, however, the
fraction of the CNO present in the SBEPs coming from the ejecta
decreases, because $\alpha_{\mathrm{ISM}}(\mathrm{CNO})$ in
Eq.~(\ref{SBEPAbundances}) scales with $Z_{\mathrm{ISM}}$.  The
contribution of the fresh SN ejecta to the total Be yield then becomes
marginal for sufficiently large values of $Z_{\mathrm{ISM}}$, and the
Be production efficiency is proportional to the ambient metallicity,
as in the standard GCRN scenario.  Indeed, the SB model is then
equivalent to the GCRN since the SBEP composition is very close to
that of the ISM. Unsurprisingly, thus, we find that the Be production
efficiency and the Be/O yield ratio is proportional to
$Z_{\mathrm{ISM}}$ for a sufficiently high ambient metallicity.

%%%%%%%%%%%%%%
\begin{figure}
\centerline{\psfig{file=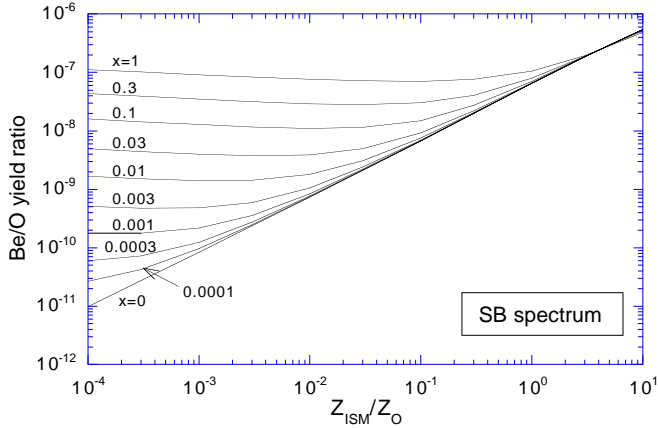,width=\columnwidth}}
\caption{Same as Fig.~\ref{Be/O(CRS2.0)}, but for the weak SB spectrum
with a cut-off energy of 500~MeV/n.}
\label{Be/O(SB500)}
\end{figure}
%%%%%%%%%%%%%%

Most importantly, the transition metallicity at which the Be evolution
changes from a primary behaviour (Be/O $\sim$ constant) to a secondary
behaviour (Be/O $\propto$ O) depends on the mixing coefficient, $x$. 
This again is evident from Eq.~(\ref{SBEPAbundances}).  The more we
put ejecta in the SBEPs (i.e. the higher the parameter $x$), the later
the secondary behaviour overcomes the primary one.  Therefore, we can
hope to predict the transition metallicity, $Z_{\mathrm{t}}$, from a
SB dynamical model (which would give the parameter $x$
self-consistently) or alternatively we can constrain the SB evolution
models from the value of $Z_{\mathrm{t}}$ derived from the
observations.  For example, Fig.~\ref{Be/O(CRS2.0)} indicates that a
transition metallicity between $[\mathrm{O}/\mathrm{H}] = -2$ and
$[\mathrm{O}/\mathrm{H}] = -1.5$ implies that the SN ejecta represent
between 1\% and 10\% of the SBEPs, which sets constraints on the SB
evolution models and/or the mixing of the gas inside SBs (see
Sect.~\ref{sec:SBmodel} and below).  However, the transition
metallicity is not the only observable which one has to reproduce.  As
discussed in Sect.~\ref{sec:energetics}, the data obtained from halo
stars imply that the Be/O yield ratio in the early Galaxy must be of
order $4\,10^{-9}$.  This is what we get from our model with the
CRS2.0 spectrum, as shown in Fig.~\ref{Be/O(CRS2.0)}, if the mixing
coefficient, $x$, is about $50\%$.  But this is out of the range
derived above.  In other words, the two constraints ($Z_{\mathrm{t}}$
and Be/O) are in contradiction with one another if we assume for the
SBEPs the CRS2.0 energy spectrum.  We should also note that a value of
$x$ as high as $\sim 0.5$ is very hard to justify since it implies
that the particles accelerated inside the superbubbles are almost pure
ejecta, in contradiction with most SB evolution models which find a
total mass inside the superbubble much larger than the mass of the
ejecta (see e.g. Parizot \& Drury 1999b, and below).

The situation is different, however, if one assumes that the particle
acceleration does not occur over the whole volume of the superbubble,
but only in those places where the ejecta of previous SNe have been
released.  But a theoretical justification and/or observational
evidence for this is still lacking.  Moreover, Fig.~\ref{Be/O(CRS2.0)}
shows that the transition metallicity corresponding to $x \sim 0.5$ is
above $[\mathrm{O}/\mathrm{H}] \sim -1$, which also appears in
contradiction with the data.  Of course, if one allows for more than
$10^{50}$~ergs to be imparted to the SBEPs per SN, all the curves on
Fig.~\ref{Be/O(CRS2.0)} can be shifted upwards and an agreement may be
recovered between the range for $x$ allowed by the Be/O constraint and
that allowed by the preliminary results relating to $Z_{\mathrm{t}}$. 
This, however, would not occur unless we increase the Be yield by a
factor of $\sim 20$, which amounts to put $2\,10^{51}$ ergs into
SBEPs, a rather unreasonable solution.

Much better appears to be the situation when the SB spectrum (either
weak or strong) are used instead of the CRS2.0 spectrum.  In
Fig.~\ref{Be/O(SB500)}, we show the results corresponding to the weak
SB spectrum with the cut-off energy $E_{0} = 500$~MeV/n.  It can be
seen that the correct Be/O yield ratio below $[\mathrm{O}/\mathrm{H}]
= -2$ is obtained for a mixing coefficient of a few percent.  This is
particularly interesting since the `standard' SB evolution models (in
the smoothly distributed ISM) predict a very similar value for the
fraction of the mass inside the bubble consisting of SN ejecta. 
Indeed, as discussed in Parizot \& Drury (1999b), the SB mass derived
from the non-magnetic evolution model (Weaver et al. 1977; Mac Low \&
McCray 1988) is given by:
\begin{equation}
	M_{\mathrm{SB}}(t) = 
	(1600~\Msun)\,L_{38}^{27/35}n_{0}^{-2/35}t_{\mathrm{Myr}}^{41/35},
	\label{MSB}
\end{equation}
where $L_{38}$ is the mechanical luminosity of the OB association
responsible for the growth of the superbubble, in units of
$10^{38}\,\mathrm{erg}\,\mathrm{s}^{-1}$, $n_{0}$ is the ambient
density in $\mathrm{cm}^{-3}$, and $t_{\mathrm{Myr}}$ is the time
elapsed since the first SN explosion (in Myr).  As a consequence, when
50 SNe have exploded, say, the time elapsed is about $t_{\mathrm{Myr}}
= 15.8$ (if $L_{38}$ = 1) and the total mass of the gas inside the SB
is $M_{\mathrm{SB}} \sim 4.1\,10^{4}\,M_{\odot}$.  Considering that
each supernova ejects, on average, $\sim 20\,M_{\odot}$ of material,
the total mass of the ejecta reaches about $1000\,M_{\odot}$, i.e.
$\sim 2.5\%$ of $M_{\mathrm{SB}}$.

That this value is precisely in the range required to account for the
Be/O ratio observed in halo stars is very remarkable, since we have
not made any particular assumption to derive it.  In other words, by
using the standard SB evolution model and the standard efficiency for
a particle acceleration model (i.e. $\sim 10\%$), one gets exactly the
amount of Be production needed to explain the Be abundances observed
in stars of metallicity lower than $10^{-2}$ solar.  This should be
considered as a strong argument in favour of both the SB model and the
SB spectra.  In addition, we see from Fig.~\ref{Be/O(SB500)} that the
very same set of parameters also leads to a transition metallicity
between $[\mathrm{O}/\mathrm{H}] = -2$ and $[\mathrm{O}/\mathrm{H}] =
-1.5$, i.e. again exactly in the range suggested by the observational
data.  Both constraints are thus found to be consistent if one uses
one of the SB spectra.  And finally, it is possible to calculate the
Be/O ratio predicted by this model at solar metallicity.  To do so, we
first need to recognize that the quantity plotted on the figures is
not the stellar Be/O abundance ratio itself, but the Be/O yield ratio
(i.e. `at production').  However, it is easy to derive one ratio from
the other: focusing on the part of the curve above $Z_{\mathrm{t}}$,
we can write
\begin{equation}
	\frac{\mathrm{Be}}{\mathrm{O}}\Big|_{\mathrm{prod}}(t) = K\times
	\mathrm{O}(t)
	\label{eq:Be/OProd}
\end{equation}
for the Be/O production ratio, where $K$ is a constant which we do not
need to specify here, and $\mathrm{O}(t)$ is the total number of O
nuclei in the ISM. Now the total number of Be nuclei in the ISM is
given as a function of time by:
\begin{equation}
\begin{split}
	\mathrm{Be}(t) &= \mathrm{Be}(t_{0}) + 
	\int_{t_{0}}^{t}\dot{\mathrm{Be}}(t^{\prime})\d t^{\prime}\\
	&= \mathrm{Be}(t_{0}) + \int_{t_{0}}^{t}\frac{\mathrm{Be}}
	{\mathrm{O}}\Big|_{\mathrm{prod}}(t^{\prime})
	\dot{\mathrm{O}}(t^{\prime})\d t^{\prime}\\
	&= \mathrm{Be}(t_{0}) +
	\int_{\mathrm{O}(t_{0})}^{\mathrm{O}(t)}K\mathrm{O}\d\mathrm{O}\\
	&= \mathrm{Be}(t_{0}) +
	\frac{K}{2}\left[\mathrm{O}^{2}(t) - \mathrm{O}^{2}(t_{0})\right],
	\label{eq:Be(t)}
\end{split}
\end{equation}
where $t_{0}$ is the time corresponding to the transition towards the
secondary behaviour of Be.  Now acknowledging that $\mathrm{Be}(t_{0})$
and $\mathrm{O}(t_{0})$ are negligible as compared to the solar
values, we obtain $\mathrm{Be}(t)\simeq\tfrac{1}{2}K\mathrm{O}^{2}(t)$
from which we deduce the elemental abundance ratio at solar
metallicity:
\begin{equation}
	\frac{\mathrm{Be}(t)}{\mathrm{O}(t)}\Big|_{\odot} = 
	\frac{1}{2}\times
	\frac{\mathrm{Be}}{\mathrm{O}}\Big|_{\mathrm{prod}}(Z_{\odot}).
	\label{eq:Be/OSolar}
\end{equation}

From Fig.~\ref{Be/O(SB500)} we see that the Be/O production ratio at
solar metallicity is $\sim 6.6\,10^{-8}$.  The `predicted' value for
the solar Be/O abundance ratio for this model is then $\sim
3.3\,10^{-8}$, remarkably close to the measured value, namely
$3.1\,10^{-8}$ (Anders \& Grevesse 1989)!  We can thus report, in
conclusion, that the SB model described above is fully consistent with
the available data, and reproduces the whole scheme of Be evolution,
both qualitatively and quantitatively, namely
\begin{enumerate}
	\item the primary behaviour of Be up to $Z_{\mathrm{t}}$,

	\item at the correct level of Be/O $\sim 4\,10^{-9}$,

	\item the secondary behaviour above $Z_{\mathrm{t}}$,
	
	\item with a value of $Z_{\mathrm{t}}$ consistent with the data,

	\item and the correct value of the Be/O abundance ratio at solar
	metallicity.
\end{enumerate}

Having shown that the SB model successfully describes the Galactic
evolution of Be over the whole range of halo star metallicities, we
now consider the case of the other light elements, Li and B, and see
whether the LiBeB data can actually be used to set constraints on the
SB dynamical evolution.

\subsection{Lithium}

%%%%%%%%%%%%%%
\begin{figure*}
\centerline{\psfig{file=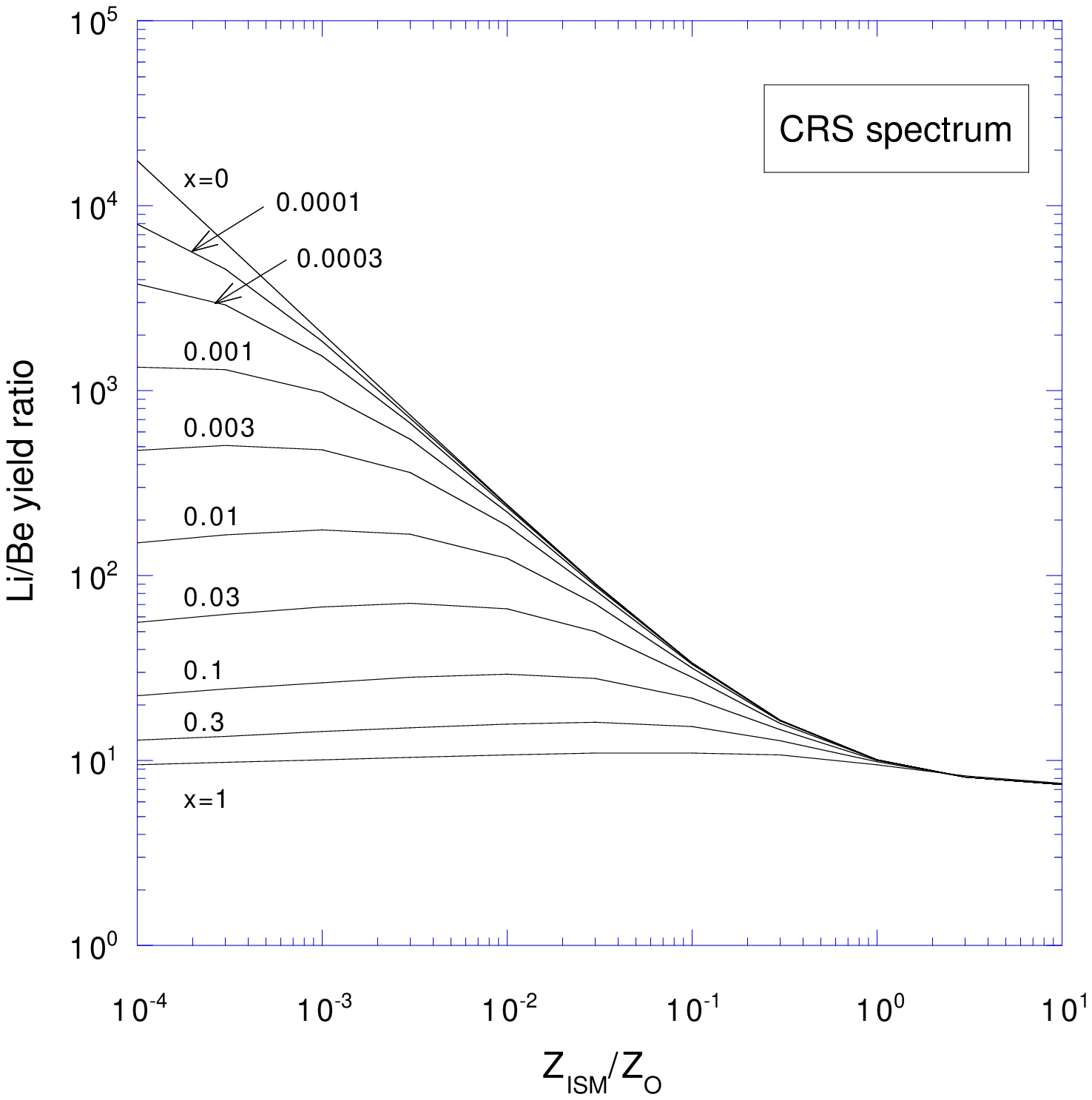,width=8.5cm}\hfill
\psfig{file=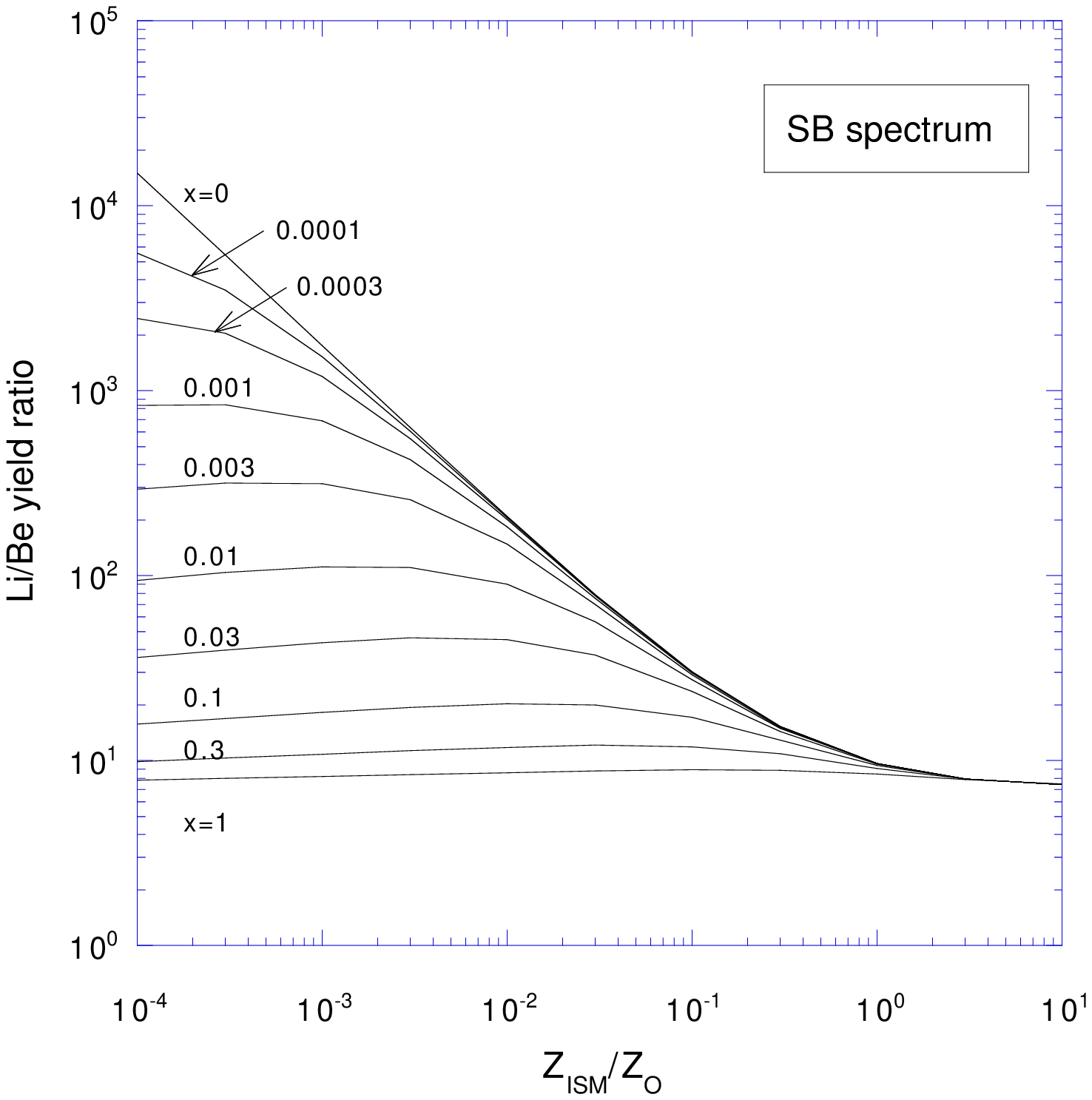,width=8.5cm}}
\caption{Li/Be yield ratio of a superbubble as a function of the
ambient metallicity, for various values of the mixing parameter, $x$. 
The SBEP spectrum is indicated on the figures.}
\label{Li/Be}
\end{figure*}
%%%%%%%%%%%%%%

The observational constraints on Li production by SBEP-induced
spallation have been summarized in Sect.~\ref{sec:phenomenology}.  In
Fig.~\ref{Li/Be}, we show the Li/Be yield ratios obtained with the
CRS2.0 and the SB spectra, for different values of the mixing
parameter, $x$.  The general shape of the curves is easily understood
if one realizes that Be is produced by CNO spallation only, while Li
is produced by CNO spallation \emph{and} $\alpha + \alpha$ reactions. 
As a consequence, higher Li/Be yield ratios are obtained for EP
compositions richer in He.  At a given metallicity, the latter
correspond to lower values of $x$, i.e. smaller proportions of
CNO-rich ejecta among the EPs.  However, as the ambient metallicity
increases, the weight of $\alpha + \alpha$ reactions decreases and the
Li production becomes dominated by the CNO-spallation anyway, so that
the spread in the Li/Be production ratios for different values of $x$
decreases.  As can be seen from the figures, the constraint
$\mathrm{Li}/\mathrm{Be} < 100$ is respected for any SB model with a
mixing parameter larger than $\sim 1\%$.  This is in very good
agreement with the range derived from the energetics constraint and
the value of $Z_{t}$ in the case of the SB spectrum.

%%%%%%%%%%%%%%
\begin{figure}
\centerline{\psfig{file=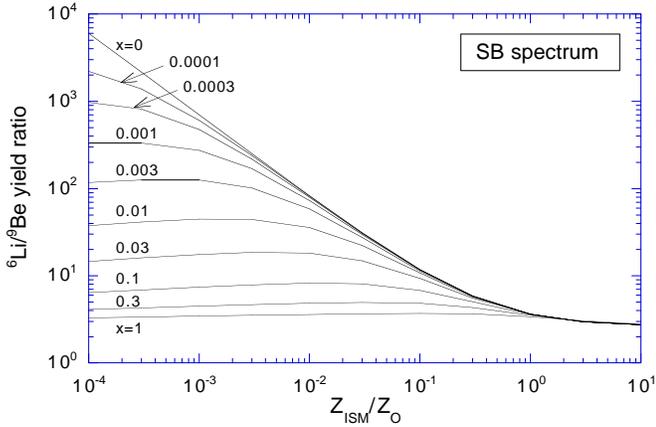,width=\columnwidth}}
\caption{$^{6}$Li/$^{9}$Be yield ratio of a superbubble as a function
of the ambient metallicity, for various values of the mixing
parameter, $x$.  The SBEP spectrum is the weak SB spectrum with a
cut-off energy of 500~MeV/n.}
\label{6Li/9Be(SB500)}
\end{figure}
%%%%%%%%%%%%%%

Concerning $^{6}$Li, we found a $^{7}$Li/$^{6}$Li isotopic production
ratio in the range 1.3--1.7 for any value of $x$, any ambient
metallicity and any SBEP spectrum.  Such a `constancy' of the isotopic
production ratio is due to the fact that $^{6}$Li and $^{7}$Li are
produced by the same nuclear reactions, with only slight variations in
the reaction thresholds and resonances.  The curves showing the
$^{6}$Li/$^{9}$Be production ratio as a function of metallicity can
thus be roughly derived from Fig.~\ref{Li/Be}, and we only show the
exact results for the case of the SB spectrum with $E_{0} =
500$~MeV/n, in Fig.~\ref{6Li/9Be(SB500)}.  Interestingly enough, the
$^{6}$Li/$^{9}$Be production ratio at solar metallicity is virtually
independent of the mixing parameter inside superbubbles, and is about
$\sim 3.6$.  In order to compare this value with the solar value,
$\sim 6$, one has to average the $^{6}$Li/$^{9}$Be production ratio
over the whole Galactic evolution.  Proceeding in the same spirit as
in Eq.~(\ref{eq:Be(t)}), we have:
\begin{equation}
	\frac{^{6}\mathrm{Li}}{^{9}\mathrm{Be}}\Big|_{\odot} =
	\frac{\int_{0}^{Z_{\odot}}\frac{^{6}\mathrm{Li}}{^{9}\mathrm{Be}}
	\big|_{\mathrm{prod}}(Z_{\mathrm{ISM}})\frac{^{9}\mathrm{Be}}{\mathrm{O}}
	\big|_{\mathrm{prod}}(Z_{\mathrm{ISM}})\d Z_{\mathrm{ISM}}}
	{\int_{0}^{Z_{\odot}}\frac{^{9}\mathrm{Be}}{\mathrm{O}}
	\big|_{\mathrm{prod}}(Z_{\mathrm{ISM}})\d Z_{\mathrm{ISM}}}\,.
	\label{eq:6/9}
\end{equation}

Performing these integrations, we obtain
$(^{6}\mathrm{Li}/^{9}\mathrm{Be})_{\odot} \sim 5$, in reasonable
agreement with the observed value.  However, as we noted above, this
constraint cannot be used to distinguish between the SB models, since
different values of $x$ and different EP spectra give roughly similar
results, namely $(^{6}\mathrm{Li}/^{9}\mathrm{Be})_{\odot}$ ratios
between 3.5 and 5.  On the other hand, Fig.~\ref{6Li/9Be(SB500)} shows
that the value of the $^{6}\mathrm{Li}/^{9}\mathrm{Be}$ ratio at lower
metallicity is a very important observable, as it allows one to
constrain the effective mixing parameter, $x$, very efficiently.  Very
remarkably, the first few data available at $\mathrm{Fe}/\mathrm{H}
\simeq -2.3$ give $^{6}\mathrm{Li}/^{9}\mathrm{Be}$ between 20 and 80,
which corresponds to a value of $x$ in a range around one percent, in
very good agreement with the values derived from the energetics and
$Z_{\mathrm{t}}$ constraints.  And once again, this agreement is
achieved very naturally with the SB spectrum, while the higher value
of $x$ required for the CRS2.0 spectrum (see above) would imply a
value of the $^{6}\mathrm{Li}/^{9}\mathrm{Be}$ ratio at low Z in
contradiction with the observation.  In this respect, we confirm that
the model proposed by Ramaty et al.  (2000a) in which the EPs have a
constant metallicity all over the Galactic chemical evolution (which
corresponds to a high $x$ in our parameterization) cannot account for
the $^{6}\mathrm{Li}/^{9}\mathrm{Be}$ ratio observed in halo stars. 
More data are expected to make this argument even stronger in the near
future (or invalidate it, if in contradiction with the current data).

In any case, it is worth stressing again that the measurement of the
$^{6}\mathrm{Li}/^{9}\mathrm{Be}$ ratio in low metallicity stars is
very important, as it allows one to constrain the SB models very
efficiently, in a way completely independent of the energetics.  And
as it stands, it is a strong argument in favour our model that all the
available constraints, although independent from one another, namely
energetics, $Z_{\mathrm{t}}$ and $^{6}\mathrm{Li}/^{9}\mathrm{Be}$,
are in excellent agreement.  Moreover, they indicate that the
proportion of the ejecta inside SBs, $x$, is of the order of a few
percent, and that the SBEP spectrum is close to the idealized SB
spectrum discussed in Sect.~\ref{SBEPEnergySpectrum}.  Both results
have strong theoretical support from unrelated fields, namely SB
dynamical evolution and SB particle acceleration (see above).

\subsection{Boron}

Finally, we turn to the case of boron.  For all the SBEP spectra and
compositions we have studied (i.e. for any value of $x$), we found
B/Be production ratios between 10 and 12, for all ambient
metallicities.  As in the case of the $^{7}$Li/$^{6}$Li production
ratio, this is due to the fact that Be and B are produced by the same
nuclear reactions, namely CNO spallation.  The same is also true for
the $^{11}$B/$^{10}$B isotopic ratio, which is found between 2.0 and
2.3 for any value of the parameters.  Comparing these values with the
observational constraints recalled in Sect.~\ref{sec:phenomenology},
we find again the well-known result that spallation processes
underproduce B, and in particular the $^{11}$B isotope, with respect
to other light elements.  As noted above, this should not be
considered as a failure of the model since other processes are known
for B production ($\nu$-process and LECR spallation) and have not been
included here.  In fact, the present results might be considered as
indications that these other processes \emph{must} be efficient in the
Galaxy.  This issue will be addressed elsewhere.

\section{Conclusions and discussion}
\label{sec:conclusion}

The SB model described above has been shown to be fully consistent
with the qualitative and quantitative constraints of LiBeB Galactic
evolution.  Notably: 1) it explains the inferred two-slope behaviour
in the framework of one single model; 2) it provides the correct value
of Be/O at low metallicity; 3) it predicts the correct value of the
transition metallicity; 4) it does not break the Spite plateau; 5) it
is consistent with the $^{6}$Li/$^{9}$Be ratio at all metallicity. 
Most importantly, these successes rely on the value of only one free
parameter, namely the proportion of the SN ejecta inside a SB. Our
calculations allow one to derive the value which best reproduces the
data, namely a few percent, from the constraints of LiBeB evolution
alone, i.e. without any external prejudice.  However, it is remarkable
that this value is exactly in the range expected from standard SB
dynamical evolution models (Mac Low \& McCray 1988).  Likewise, the
SB model is found to be successful only if the EPs accelerated inside
a superbubble have an energy spectrum flattened at low energy (in
$E^{-1}$).  Now this is just what Bykov's SB acceleration model
predicts.

Interestingly enough, the consistency of the SB model is such that one
may be tempted to reverse the argument and consider that the results
of LiBeB Galactic evolution lend support to the current ideas about SB
dynamical evolution and particle acceleration.  Silich et al.  (1996)
have studied in detail the effect of the evaporation of the clouds
engulfed inside superbubbles on their dynamical evolution.  They
assume that the ambient ISM is composed of small diffuse clouds with
typical internal density of $10\,\mathrm{cm}^{-1}$, and consider both
cloud evaporation and dynamical disruption through internal flows
inside the bubble.  In the present context, these mechanisms would
imply a modification of the mass loading of the superbubble and thus
affect the expected proportion of the ejecta among the SBEPs. 
Although our results indirectly confirm the relevance of superbubble
simulations in a smoothly distributed ISM, with only large-scale
density gradients, the models taking into account a cloudy ISM cannot
be rejected.  In particular, Bykov (1999) has shown that in
time-dependent models most of the acceleration occurs at relatively
early times, when the contribution of cloud disruption and evaporation
to the superbubble mass loading is small, according to the
calculations of Silich et al.  (1996).

As far as the particle acceleration inside superbubbles is concerned,
we find that only an energy spectrum showing the characteristic shape
of multiple shock acceleration at low energy ($Q(E)\propto E^{-1}$) is
compatible with all the data.  On the other hand, any SB spectrum is
found to give very similar results.  This means that LiBeB evolution
alone cannot be used to determine whether the SBEPs extend to high
energy or not.  Above the so-called break energy, their spectrum could
either match the standard CRS spectrum or decrease more steeply around
an energy of a few hundreds or thousands of MeV/n.  In the latter
case, the SBEPs will have to be regarded as a second component of EPs
in the Galaxy, in addition to cosmic-rays.  In the opposite case,
however, the SBEPs could be nothing but the GCRs and the SB model
could then be thought of as a mere correction of standard GCRN, taking
into account the chemical inhomogeneity of the early Galaxy.  Indeed,
GCRN assumes that the LiBeB-producing energetic particles are
accelerated out of the average ISM. In a more detailed analysis,
however, one has to acknowledge that the metals are released in
localized regions of the Galaxy and abandon the homogeneous, one-zone
representation.  For it turns out that particle acceleration occurs
precisely where the metals are more abundant.  This is the essence of
the SB model.  It makes a crucial difference in the early Galaxy, when
the chemical inhomogeneities are sharper (typically for $Z <
Z_{\mathrm{t}}$).  But the value of the parameter $x$ which we derived
(a few percent) is small enough for the chemical composition to be
roughly identical inside and outside superbubbles today.  This shows
that the SB model and the GCRN model are essentially the same at
metallicities $Z > Z_{\mathrm{t}}$, i.e. for most of the lifetime of
the Galaxy.

If the bottom line of the SB model is to account for chemical
inhomogeneities, then it could actually apply to a wider context than
SBs themselves.  Localized star bursts would behave in about the same
way, and globular clusters might also be considered as large OB
associations giving rise to the same kind of phenomena, and thus
participating to the LiBeB evolution in about the same way.  The
framework of LiBeB evolution described here would thus be quite
general and only slight variations of the relevant parameters would
occur from one environment to an other, or from one place to another. 
These variations would cause possibly strong variation of the light
elements--to--metals ratios from star to star, but all fitting in the
same general pattern, namely a primary behaviour of $^{6}$LiBeB up to
a transition metallicity, $Z_{\mathrm{t}}$, around $10^{-2}$ to
$10^{-1}\,Z_{\odot}$, and a secondary behaviour afterwards.  The
resulting scatter in the LiBeB data has been discussed in Parizot \&
Drury (2000) (see also Cunha et al., 2000).  In the spirit of the
present paper, the scatter in the Be and B data can be evaluated by
letting $x$ vary from one SB to another, or as a function of time. 
This amounts to allow for some `diffusion' across the curves
represented in Figs.~\ref{Be/O(CRS2.0)}, \ref{Be/O(SB500)} and
\ref{6Li/9Be(SB500)}.  As an example, a variation of $x$ by a factor
of 3 (from 1\% to 3\%, say) would result in a variation of Be/O by the
same factor at very low $Z$, while only by a factor of 2 at
$\mathrm{[O/H]} = -2$, and no variation at $Z = Z_{\odot}$.

Hopefully, improvements of the theoretical modeling of stellar
atmospheres and the line formation will allow one to shorten the
errors bars and measure the scatter in the data as a function of
metallicity.  Stronger constraints will thus be set on the LiBeB
evolution models.  Likewise, an accurate determination of the mass
fraction of ejecta inside the SB can only be achieved if the
transition metallicity, $Z_{\mathrm{t}}$, is measured with enough
precision.  This requires data points at metallicities much lower than
$\mathrm{[O/H]} = -2$.  As for the determination of the SBEP energy
spectrum at high energy, it shall be more efficiently constrained by
gamma-ray astronomy.  An important challenge for the satellite
INTEGRAL, to be launched in 2002, will be to detect nuclear
de-excitation lines from EP interactions in nearby superbubbles (e.g.
Orion or Vela).  The expected gamma-ray fluxes have been estimated to
be comparable with the INTEGRAL thresholds (Parizot \& Kn\"odlseder,
1999a,b).

Finally, we wish to stress one of the important differences between
the SB model, which accounts for the `two-slope behaviour' of Be and B
evolution within a single model, and an alternative scenario in which
the low-metallicity (primary) part of the correlation would be
attributed to one mechanism, and the secondary part to another
mechanism (e.g. GCRN).  In our model, the transition between the slope
1 and slope 2 Be-O correlation is continuous.  Any intermediate value
is reached over a given range of stellar metallicity (depending on the
value of $x$).  In the other case, by contrast, one would have a sharp
change of slope at the precise metallicity where the secondary process
becomes dominant, with no intermediate values.  Of course, expected
physical fluctuations of the parameters would weaken this effect, and
current observational error bars prevent us from distinguishing
conclusively between the two pictures.  But we argue that the observed
`slope 1.5' behaviour reported by some authors (e.g. Boesgaard \&
Ryan, 2000) can be explained (in principle) only if there is a
continuous \emph{transition} from slope 1 to slope 2 within a
\emph{unique} model, rather than two unrelated models with a slope~2
eventually superseding a slope~1.

\begin{acknowledgements}
I wish to thank the International Astronomical Union and the
organizers of the IAU 198 symposium on LiBeB evolution in Natal
(Brazil), which stimulated this work.  I was supported by the TMR
programme of the European Union under contract FMRX-CT98-0168.
\end{acknowledgements}

\end{document}